\documentclass[bimj,fleqn]{w-art}
\usepackage{times}
\usepackage{w-thm}
\usepackage[authoryear]{natbib}
\usepackage{url}
\theoremstyle{plain}

\theoremstyle{definition}

\usepackage[]{graphicx}
\chardef\bslash=`\\ 

\usepackage{hyperref}
\usepackage{rotating}
\RequirePackage{booktabs}%

\usepackage{etoolbox}
\AtBeginEnvironment{tabular*}{\fontsize{8}{10}\selectfont}
\AtBeginEnvironment{tabular}{\fontsize{8}{10}\selectfont}
\AtBeginEnvironment{tablenotes}{\fontsize{7}{9}\selectfont}

\newenvironment{tablenotes}{\list{}{\setlength{\labelsep}{0pt}%
\setlength{\labelwidth}{0pt}%
\setlength{\leftmargin}{0pt}%
\setlength{\rightmargin}{0pt}%
\setlength{\topsep}{2pt}%
\setlength{\itemsep}{0pt}%
\setlength{\partopsep}{0pt}%
\setlength{\listparindent}{0em}%
\setlength{\parsep}{0pt}}%
\item\relax%
}{\endlist}%

\def\tnote#1{#1}

\begin{document}
\Volume{00}
\Issue{00}
\Year{2025}
\pagespan{1}{}
\keywords{Corrected Likelihood; Covariate Measurement Error; Poisson Surrogate; POI-SIMEX; Tissue Microarrays\\
\noindent \hspace*{-4pc} {\small\it }\\
\hspace*{-4pc} {\small\it }\\[1pc]
}  

\title[POI-SIMEX]{POI-SIMEX for Conditionally Poisson Distributed Biomarkers from Tissue Histology {\small\it }}
\author[Yang A. {\it{et al.}}] {Aijun Yang \inst{1}}
\address[\inst{1}]{3800 Finnerty Rd {\it(Department of Mathematics \& Statistics,University of Victoria, V8P5C2, BC, Canada)}}
\author[]{Phineas T. Hamilton\inst{2}}
\address[\inst{2}]{2410 Lee Avenue, Victoria {\it{(Deeley Research Center, BC Cancer Agency, V8R6V5, BC, Canada)}}}
\author[]{Brad H. Nelson\inst{2,3}} {\small\it}
\address[\inst{3}]{3800 Finnerty Rd {\it(Department of Biochemistry and Microbiology,University of Victoria, V8P5C2, BC, Canada)}}
\author[]{Julian J. Lum\inst{2,3}} {\small\it}
\author[]{\\Mary Lesperance\inst{1}} {\small\it}
\author[]{Farouk S. Nathoo \footnote{Corresponding author: Farouk S. Nathoo {\sf{e-mail: nathoo@uvic.ca}}}\inst{,1}} {\small\it}


\begin{abstract}
Covariate measurement error in regression analysis is an important issue that has been studied extensively under the classical additive and the Berkson error models. Here, we consider cases where covariates are derived from tumor tissue histology, and in particular tissue microarrays. In such settings, biomarkers are evaluated from tissue cores that are subsampled from a larger tissue section so that these biomarkers are only estimates of the true cell densities. The resulting measurement error is non-negligible but is seldom accounted for in the analysis of cancer studies involving tissue microarrays.

To adjust for this type of measurement error, we assume that these discrete-valued biomarkers are conditionally Poisson distributed, based on a Poisson process model governing the spatial locations of marker-positive cells. Existing methods for addressing conditional Poisson surrogates, particularly in the absence of internal validation data, are limited. We extend the simulation extrapolation (SIMEX) algorithm to accommodate the conditional Poisson case (POI-SIMEX), where measurement errors are non-Gaussian with heteroscedastic variance. 

Simulation studies evaluate the performance of POI-SIMEX, comparing it with the naive method and an alternative corrected likelihood approach in linear regression and survival analysis. POI-SIMEX is then applied to a study of high-grade serous ovarian cancer, examining the association between survival and the density of CD3+CD8+FOXP3+ immune cells.
\end{abstract}

\maketitle                   






\section{Introduction} \label{sec1}
Accurate covariate measurement is critical in clinical studies, as error can severely bias estimates and undermine prognostic models \citep{Carroll2006,Fuller1986}. While \citet{Yang2025} and its references provide a broad overview of recent methodological advances for covariate measurement error in linear regression settings, a focused review within survival analysis reveals a diverse toolkit. Numerous methods have been adapted for various models and assumptions \citep{Carroll2006}. For the Cox proportional hazards model, common techniques include the corrected score approach introduced by \cite{Nakamura1992} and later refined \citep{Kongetal1998,KongGu1999,Augustin2004,Yi2007,Yan2015}, and the approximate profile likelihood method of \cite{Cao2022}. The regression calibration method offers a straightforward solution when validation data are available \citep{Spiegelman1997}. For parametric accelerated failure time (AFT) models, the simulation-extrapolation (SIMEX) technique has been effectively applied \citep{He2007AFT,Zhang2014AFT,Greene2004AFT}. This approach facilitates the modeling of flexible, non-linear covariate effects within a AFT framework, utilizing a generalized additive model structure. Coupled with a classical measurement error model, the methodology enables rigorous investigation of the asymptotic properties of the associated SIMEX estimator.
Bayesian methods incorporate prior information and better allow for uncertainty quantification with finite sample inference. A Bayesian approach to covariate measurement error can be formulated directly by treating the true covariates as missing data and using data augmentation \citep{Richardson1993}. On the other hand, nonparametric regression techniques \citep{Carroll1999,Carroll2009}, like kernel regression, spline smoothing, and Dirichlet process mixtures \citep{Sarkar2014,Sinha2017}, offer flexibility without assuming specific functional or distributional forms. 

Our methodological development is motivated by studies investigating the prognostic significance of tumor-infiltrating lymphocytes with TMAs, and our example application considers this in a study of high-grade serous ovarian cancer (HGSC). HGSC remains the deadliest gynecological malignancy primarily because of its early spread and the development of resistance to platinum-based chemotherapy as mentioned in \cite{Havasi2023}. Thus, there is a critical need for reliable biomarkers that complement clinical data to support advances in patient care. These biomarkers are typically measured through sub-sampling a single disease site; therefore, they are prone to measurement errors arising from TMA construction, and leading to non-Gaussian, heteroscedastic measurement errors associated with discrete biomarkers. While the corrected score estimator for Poisson surrogates proposed by \cite{Lietal2004} can be applied in this context, its application is limited to linear regression. In contrast, SIMEX can be applied more broadly, but it has yet to be developed for conditionally Poisson-distributed surrogates. In addition, even in the context of linear regression, we have found that SIMEX generally outperforms the corrected score methodology.

To address this, we recently established conditions under which SIMEX is strongly consistent in linear regression as well as an approach to consistently estimate the measurement error variance \citep{Yang2025}. POI-SIMEX, a novel extension of the simulation-extrapolation (SIMEX) method is  tailored for regression settings with covariate measurement error where observed surrogates are conditionally Poisson-distributed. This paper moves beyond the theoretical foundation to focus on its practical application to cancer survival study involving tissue histology. Our primary contributions are the development and comprehensive evaluation of the POI-SIMEX algorithm, providing a much-needed tool for data analysis in TMA-based cancer research and other fields dealing with similar measurement error structures.

Subsequent sections of the paper proceed as follows. Section 2 presents an introduction to TMA-based biomarkers and describes the mechanisms that lead to measurement error. Section 3 focuses on the estimation bias that arises if measurement error is ignored under a conditional Poisson error model and the POI-SIMEX methodology. Section 4 presents extensive simulation studies that demonstrate its superior finite-sample performance, followed by an application to HGSC in Section 5. The paper concludes with a discussion of our results, including the relevance to both statistical methodology and cancer research.

\section{Tissue Microarrays and Cancer Biomarkers}\label {sec2}

Patients diagnosed with cancer have traditionally received standardized treatment protocols, however, it is now clear that unique characteristics of a patient’s tumor can guide tailored therapies even among cases of the same cancer type. Beyond clinically approved biomarkers, such as the expression of PD-L1 by cancer and tumor-infiltrating immune cells, biomarker characterization has become increasingly important to guide research and clinical development. Many studies investigate associations between certain cell types and patient outcomes. At the same time, advances in digital pathology are leading to increasingly quantitative characterization of tissue biomarkers that go beyond binary positive or negative status for a given marker. The consideration of measurement error for these biomarkers is of great importance in the analysis of such data.

Tissue microarrays are widely used in cancer research because they enable the high-throughput analysis of multiple tissue samples on a single slide, significantly increasing research efficiency and reducing costs. TMAs ensure consistency and standardization in sample processing, enhancing reliability and reproducibility. They also conserve valuable tissue specimens by using small cores and allow for comprehensive biomarker analysis, facilitating the validation of known markers, the discovery of new markers and the analysis of marker interaction. Their ability to utilize archival tissue further supports their extensive application in cancer research \citep{Kononen1998}. 

\begin{figure*}
\centerline{\includegraphics[scale=0.95]{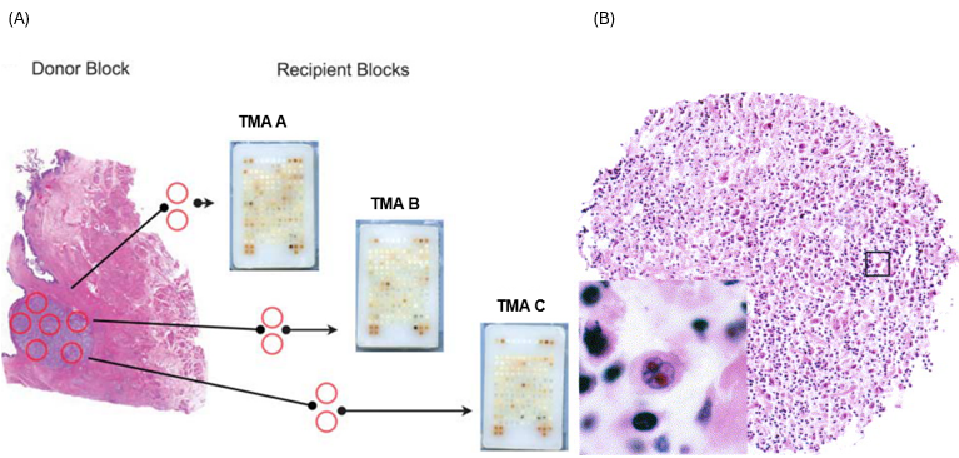}}
\caption{
(A): Tissue microarray construction. TMA Cores are extracted from donor blocks using an automated arrayer guided by the annotated digital slides. Slides containing the desired target tissue are digitally scanned and annotated using the 3D Histotech system. (image credit \citep{Eskaros2017});
(B) Tissue core of a Hodgkin's lymphoma tumor. Inset: Magnification of the framed region where the ``owl's eye" shape is a Reed-Sternberg cell (image credit \citep{Ronald2002}).\label{fig1}}
\end{figure*}

Tissue cores are extracted by puncturing solid tissue using a circular needle typically ranging from 0.6 mm to 4.0 mm in diameter, depending on tissue availability, resulting in only small subregions of tumors that are sampled as cores. Consequently, a TMA will only approximate the cell density of interest \cite{TMAIntro}. Depending on the study, replicate cores may or may not be available for each subject. Methods that do not require replication are thus more generally useful. Importantly, these measures are discrete rather than continuous as they are computed based on cell counts. The resulting measurement error is non-negligible and the discrete nature of the observed covariate makes this a challenging problem that is rarely considered in the analysis of TMAs. Figure \ref{fig1} (A) illustrates a TMA sample construction \citep{Eskaros2017}. In this case, a TMA core (red circle) with area $A$ and biomarker cell count $W$ is extracted from a whole donor tissue (denoted as T) with an area $A_T$ and a total biomarker cell count $W_T$, where $A << A_T$. Inaccuracies can arise when analyzing marker expression based on these core counts and core areas leading to bias and inflated variance in regression analysis when TMA-based biomarkers are incorporated into a regression analysis.

\section{Methodology} \label{sec3}
\subsection{Poisson error model assumptions}
We model $Y$, the response variable, as a function of $X$, the true biomarker of interest, where $X$ typically represents cell density $W_T/A_T$ over a larger tissue area, which cannot be measured directly with TMAs. Instead, the variable $W$, representing the biomarker cell count of the selected TMA core with area $A$ is measured. The estimated or observed cell density is $W/A$ and is used as the surrogate for $W_T/A_T$. Other covariates $Z$ are assumed to be measured without error. Cell locations within a tissue sample are assumed to be distributed according to a homogeneous Poisson process with intensity $X = \frac{W_T}{A_T}$. 
Under the Poisson process model for the cell distribution we have for subject $i$,
\begin{equation}
\label{eq0}
W_i\lvert X_i \overset{\displaystyle \text{ind}}{\sim} \text{Poisson} (X_i A_i) \ \ i=1,...,N, 
\end{equation}
where $X_i$ is the unobserved true tissue cell density for the $i^{th}$ subject and $W_i$ is the observed number of cells on the core with area $A_i$. Thus, $E[W_i\lvert X_i]= Var [W_i\lvert X_i]=A_i X_i$, and $E[W_i]=A_i E[X_i]$. We note that the model is Poisson conditional on $X_{i}$, and therefore overdispersion in the observed $W_{i}$ does not preclude the use of this model. 

In what follows, we assume $(Y_i,Z_i,W_i,A_i,X_i)$, corresponding to the response, covariate without measurement error, observed cell count, observed core area, and true cell density, respectively, are independent across subjects $i=1,...N$. For simplicity, we illustrate the conditional Poisson measurement error model in the context of a linear model and assume both $X_i$ and $Z_i$ are scalars. We then have
\begin{equation}
Y_i =\beta_0 + \beta_XX_i + \beta_ZZ_i + \epsilon_i, \ \ \epsilon_i \overset{\displaystyle \text{i.i.d}}{\sim} \text{N} (0,\sigma_\epsilon^2) , \ \  W_i\lvert X_i \overset{\displaystyle \text{ind}}{\sim} \text{Poisson} (X_i A_i), 
\label{eq1}
\end{equation}
where $X_i >0$ is the true cell density, and  $(Y_i,Z_i,W_i,A_i)$ $i=1,..., N$ are the observed data.  We first discuss asymptotic bias for the case when $X_i$ is replaced by $W_i/A_i$, the so-called naive estimator.

\subsection{The Asymptotic bias of the naive estimator} \label{subAsyBias}
The bias analysis associated with the conditionally Poisson distributed surrogate follows that in Li \textit{et al.} \citeyearpar{Lietal2004}, though we provide more detail, which clarify the results. 
\noindent Letting $Q^t = (\frac{W_1}{A_1},...,\frac{W_N}{A_N}), \Lambda=(X_1,...,X_N)^t$, $M^t=\left(\overset{\displaystyle \text{ 1 1 ...1 }}{Z_1 Z_2 ... Z_N}\right) $, 
the least squares estimator for the regression coefficient associated with the naive model, $Y_i = \beta_0+\beta_X(W_i/A_i) + \beta_ZZ_i + \epsilon_i^*$, $Y=(Y_1,...,Y_N)^t$, $\epsilon_i^* \overset{\displaystyle \text{i.i.d}}{\sim} N(0, \sigma_{\epsilon^*}^2)$, when 
(\ref{eq1})  is the true model, can be written as in Li \textit{et al.} \citeyearpar{Lietal2004}  
\begin{align*} 
\hat{\beta_X}^{(N)}=\frac{\left[Q^t-Q^tM(M^tM)^{-1}M^t\right ]Y}{Q^tQ-Q^tM(M^tM)^{-1}M^tQ}  \text{ ,}
\end{align*}
from which we obtain the conditional expectation 
\begin{align*} 
E\left[\hat{\beta_X}^{(N)}\lvert M,Q,\Lambda\right]=\frac{\left[Q^t-Q^tM(M^tM)^{-1}M^t\right]\Lambda}{Q^tQ-Q^tM(M^tM)^{-1}M^tQ} \beta_X \text{ .}
\end{align*}
Taking the limit of this conditional expectation
\begin{align*} 
\lim_{N\to\infty}\frac{\left[Q^t-Q^tM(M^tM)^{-1}M^t\right]\Lambda}{Q^tQ-Q^tM(M^tM)^{-1}M^tQ} \beta_X
=\lim_{N\to\infty} \frac{\left[\frac{1}{N}Q^t \Lambda-\frac{1}{N}Q^tM(\frac{1}{N}M^tM)^{-1}\frac{1}{N}M^t\Lambda\right]}{\frac{1}{N}Q^tQ-\frac{1}{N}Q^tM(\frac{1}{N}M^tM)^{-1}\frac{1}{N}M^tQ} \beta_X \text{ .}
\end{align*}
Taking the limit of each term on the right-hand side:

\noindent i) 
\vspace{-1em}
\begin{align*} 
\lim_{N\to\infty} \frac{1}{N}Q^t\Lambda=\lim_{N\to\infty} \frac{1}{N}\sum_{i=1}^{N}\frac{W_i}{A_i}X_i \overset{p}{\rightarrow} \ \ E\left[\frac{W_i}{A_i}X_i\right] =E\left\{E\left[\frac{W_i}{A_i}X_i \lvert X_i\right]\right\}=E\left[X_i^2\right]
\vspace{-2em}
\end{align*}
\noindent ii) 
\vspace{-2em}
\begin{align*} 
\lim_{N\to\infty} \frac{1}{N}Q^tQ &=\lim_{N\to\infty} \frac{1}{N}\sum_{i=1}^{N}\frac{W_i^2}{A_i^2}\overset{p}{\rightarrow} 
 \ \ E\left[\frac{W_i^2}{A_i^2}\right] =\frac{1}{A_i^2}E\left\{E\left[W_i^2 \lvert X_i\right]\right\}\\
&=\frac{1}{A_i^2}E\left[X_iA_i+X_i^2A_i^2\right] =\frac{E\left[X_i\right]}{A_i} + E\left[X_i^2\right]
\end{align*}

\noindent iii) 
\vspace{-2em}
\begin{align*} 
\lim_{N\to\infty} \frac{1}{N}Q^tM&=\lim_{N\to\infty} \left ( \frac{1}{N}\sum_{i=1}^{N}\frac{W_i}{A_i}, \frac{1}{N} \sum_{i=1}^{N} \frac{W_iZ_i}{A_i} \right )
\overset{p}{\rightarrow} \ \ \left (E\left[\frac{W_i}{A_i}\right],E\left[\frac{W_iZ_i}{A_i}\right] \right )\\
&=\left (E\left\{E\left[\frac{W_i}{A_i} \lvert X_i\right]\right\}, E\left\{E\left[\frac{W_iZ_i}{A_i} \lvert X_i\right] \right\} \right )\\
&=\left( E\left[X_i\right],\frac{1}{A_i} E\left\{E\left[W_i \lvert X_i\right]E\left[Z_i \lvert X_i\right]\right\}\right )\\
&=\left (E\left[X_i\right], E\left\{X_i E\left[Z_i \lvert X_i\right]\right\} \right ) =E\left\{X_i \left (1,E\left[Z_i \lvert X_i\right]\right)\right\} \\
&=E\left\{X_i \left (E\left[1 \lvert X_i\right],E\left[Z_i \lvert X_i\right]\right )\right\}=E\left\{X_iE\left[(1,Z_i)\lvert X_i\right]\right\} \\
&=E\left\{X_iE\left[M_i^t\lvert X_i\right]\right\}=E\left\{E\left[X_iM_i^t\lvert X_i\right]\right\}\\
&=E\left[X_i M_i\right]^t \text{ where } M_i \text{ is the } i^{\text{\tiny th}} \text{ column of } M
\end{align*}

\noindent iv)
\vspace{-2em}
\begin{align*} 
\lim_{N\to\infty} \frac{1}{N}M^t\Lambda=\lim_{N\to\infty}\left(
  \begin{matrix}
    \frac{1}{N}\sum_{i=1}^{N} X_i\\
    \frac{1}{N}\sum_{i=1}^{N} Z_iX_i
  \end{matrix}
  \right)
= \left(\begin{matrix}
    E[X_i]\\
    E[Z_iX_i]
  \end{matrix}
  \right)
=E[X_iM_i]
\end{align*}
v)
\vspace{-2em}
\begin{align*} 
\lim_{N\to\infty} \frac{1}{N}M^tM=E[M_iM_i^t]
\end{align*}
Putting (i) to (v) together we obtain
\begin{align*} 
\lim_{N\to\infty}E[\hat{\beta_X}^{(N)} \lvert M,Q,\Lambda]&=\frac{E[X_i^2]-E[X_iM_i]^t (E[M_iM_i^t])^{-1}E[X_iM_i]}{E[X_i^2]+E[\frac{X_i}{A_i}] -E[X_iM_i]^t (E[M_iM_i^t])^{-1}E[X_iM_i]} \beta_X\\
&=\omega \beta_X,\text{ where } 0<\omega <1.
\end{align*}
Thus, analogous to the case of the additive measurement error model, we see the asymptotic bias of the naive estimator in the case of a conditional Poisson model for the observed covariate in a linear regression analysis.

\subsection{POI-SIMEX methodology}\label{subPoiSimex} 
Introduced by \cite{Cook1994}, the SIMEX algorithm is widely used to mitigate bias arising from measurement error in regression models, relying on specific assumptions about the underlying error structure. The POI-SIMEX method is an extension of the SIMEX algorithm designed to correct for measurement error when the observed surrogate covariate followed a conditional Poisson distribution as shown in Eq.(\ref{eq0}). The POI-SIMEX framework diverges from the conventional SIMEX methodology, which assumes an additive structure with errors distributed as $N(0,\sigma^2)$, where $\sigma^2$ is either known or estimated. The standard SIMEX procedure comprises three sequential phases: simulation, estimation, and extrapolation. During the simulation phase, synthetic measurement error is incrementally introduced into the observed data. In the estimation phase, model parameters are repeatedly estimated across varying levels of induced error. The extrapolation phase then examines the relationship between error magnitude and parameter estimates, enabling projection of the estimators to the hypothetical case of no measurement error. The implementation of the algorithm requires estimation of the subject-specific measurement error variance, $\sigma_i^2$, for $i=1,\dots, N$. A key aspect of our proposed POI-SIMEX and an important issue for TMA data is consistent estimation of the heteroscedastic measurement error variance from the observed data with only a single TMA replicate. From Proposition 2.1 of \cite{Yang2025}, a consistent estimate of $\sigma_i^2$, the measurement error variance, can be obtained without any covariate replication.

\subsection{Variance estimation of the SIMEX estimator} \label{subVarEstSimex} 
The variance ($\sigma_{\beta_{SIMEX}}^2$) of the SIMEX estimator is comprised of two components: the Monte Carlo sampling variance of the $\hat{\beta}_\lambda$ estimates obtained at each $\lambda$ value, denoted as $\hat{\eta}_{\beta}^2(\lambda)$, and the variance of $\hat{\beta}_\lambda$ derived using the information matrix at each level of $\lambda$, denoted as $\hat{\sigma}_\beta^2(\lambda)$. \cite{Carroll2006} suggest computing this variance by extrapolating the difference of $\hat{\sigma}_\beta^2(\lambda)-\hat{\eta}_{\beta}^2(\lambda)$ to $\lambda = -1$. We illustrate this procedure in the one covariate case. In a simulation step, for a given $\lambda$ and the $b^{th}$ simulated dataset ($b=1,..,B$), let $\hat{\beta}_{\lambda,b}$ and $\hat{\sigma}_\beta^2(\lambda,b)$ be the naive estimator of the $b^{th}$ simulation for a given $\lambda$ and its corresponding variance estimator. We obtain $B$ estimates of $\hat{\beta}_{\lambda,b}$ and $\hat{\sigma}_\beta^2(\lambda,b)$ from B simulated datasets. Then, we calculate $\hat{\sigma}_\beta^2(\lambda)$ by $\frac{\sum_{b=1}^B\hat{\sigma}_\beta^2(\lambda,b)}{B}$, and $\hat{\eta}_{\beta}^2(\lambda)$ is the sample variance of the $\hat{\beta}_{\lambda,b}, b=1,\dots,B$ across simulations. We apply extrapolation to the fitted function of positive $\hat{\sigma}_{diff}^2(\lambda)=\hat{\sigma}_\beta^2(\lambda)-\hat{\eta}_{\beta}^2(\lambda)$ versus $\lambda$ values. The estimated variance of the SIMEX estimator is then $\hat{\sigma}_{\beta_{SIMEX}}^2=\hat{\sigma}_{diff}^2(\lambda=-1) $.

Another approach to estimating the variance of the SIMEX estimator is proposed by \cite{Lin2000}, who employ the sandwich formula for variance estimation of the naive estimate. This method then adjusts the result by multiplying it with a factor incorporating the extrapolation function to accommodate for it. It is well known that the bootstrap \citep{Efron1979} method can estimate the variance robustly. The variance estimate from the first approach is compared to that obtained using the nonparametric bootstrap in our application. 

\section{Simulation studies} \label {sec4} 
Simulation studies are carried out to empirically evaluate the bias and mean squared error (MSE) of POI-SIMEX, the estimator obtained using the true covariate, the Naive estimator, and the Adj-LM approach proposed by \cite{Lietal2004}. We use a consistent estimator of the measurement error variance when implementing POI-SIMEX and adopt a quadratic extrapolant function \citep{Cook1994}. An extensive range of settings are considered, including those considered in \cite{Lietal2004} and various other settings based on ratios of the variance of the true covariate and the observed covariate. These ratios inversely represent the relative magnitude of the measurement error in the observed covariate. 

\subsection{Simulation settings in Li et al.}\label{subsim} 
The data are simulated assuming the linear regression model (\ref{eq1}) and the observed count $W$ in an area $A$ follows a Poisson distribution with the mean defined as the product of the true covariate ($X$) and the area ($A$). Hence, the surrogate for the true covariate is $W/A$ where $W\lvert X \sim Poisson(XA)$. For simplicity, the area $A$ is set to 1 in the simulations. In addition to the covariate with measurement error, there is another covariate $Z$ measured without error generated from a uniform distribution between 0.5 and 9. The coefficients associated with the intercept,  covariate with measurement error and without measurement error, respectively $\beta_0$, $\beta_X$, $\beta_Z$, are set to 2, 1, and 0.5, following \cite{Lietal2004}. The sample size N is set to be 50, 100, or 200. The standard deviation of the error ($\sigma_\epsilon$) in the regression model is set to 5. Three scenarios are considered: $X_i$’s are generated independently from a Gamma distribution with shape parameter, a, and scale parameter, b as follows: (1) a = 1 and b = 2; (2) a = 1 and b =10; (3) a = 2 and b = $Z_i$. In the third case, the two covariates, $X$ and Z, are dependent, as one of the parameters used to simulate $X_i$ is set to $Z_i$. 

In all instances, the bias and mean squared error (MSE), along with their associated Monte Carlo (MC) standard errors, are reported from 1000 simulated data sets using the batch mean method with ten batches. The results for all estimated $\beta's$ and methods are summarized in Tables~\ref{tab1},~\ref{tab2}, and ~\ref{tab3}. In all settings, POI-SIMEX and Adj-LM outperform the Naive method. In scenarios 1 and 2, POI-SIMEX consistently outperforms Adj-LM, exhibiting lower MSE and bias across all settings. In scenario 3, POI-SIMEX performs better than Adj-LM with smaller MSE and bias for N=50 and N=100. For N=200, POI-SIMEX has a smaller MSE but a slightly larger bias than Adj-LM. Overall, POI-SIMEX exhibits improved performance for dealing with Poisson surrogates in linear regression. However, perhaps the greatest advantage of POI-SIMEX over Adj-LM is that it can be easily adapted to settings with censoring or non-linear regression settings while Adj-LM cannot and is based explicitly on linear regression. This is demonstrated in Section \ref{subsim_lognorm}.
\begin{sidewaystable}
\def\d{\hphantom{0}}
\caption{Simulation results for scenario 1: $X_i \sim Gamma(1,2)$, $W_i \lvert X_i \sim Poisson (X_i)$ and sample sizes 50, 100, and 200 \label{tab1}}
\begin{tabular*}{\textwidth}{@{\extracolsep\fill}lllllllllll@{}}
\toprule
& &\multicolumn{3}{@{}l}{\textbf{N=50$^{\tnote{\bf *}}$}} & \multicolumn{3}{@{}l}{\textbf{N=100$^{\tnote{\bf *}}$}} & \multicolumn{3}{@{}l}{\textbf{N=200$^{\tnote{\bf *}}$}} \\\cmidrule{3-5}\cmidrule{6-8}\cmidrule{9-11}
\textbf{Parameter}& \textbf{Method$^{\tnote{\bf **}}$}  & \textbf{Estimate} & \textbf{MSE(MCSE)} & \textbf{Bias(MCSE)} & \textbf{Estimate} & \textbf{MSE(MCSE)} & \textbf{Bias(MCSE)} & \textbf{Estimate} & \textbf{MSE(MCSE)} & \textbf{Bias(MCSE)}\\
\midrule
$\beta_0$=2   & Adj-LM   & 1.7814 & 3.9389(0.1607) & -0.2186(0.0310) & 1.8514  & 1.8773(0.1110) & -0.1486(0.0404) & 1.9750  & 0.9396(0.0582) & -0.0250(0.0285) \\
          & Naive-LM & 2.6465 & 3.1129(0.1135) & 0.6465(0.0365)  & 2.6228  & 1.8252(0.0920) & 0.6228(0.0342)  & 2.6798  & 1.2412(0.0467) & 0.6798(0.0221)  \\
          & POI-SIMEX    & 2.1732 & 3.1792(0.1287) & 0.1732(0.0347)  & 2.1455  & 1.6623(0.0937) & 0.1455(0.0381)  & 2.2159  & 0.9086(0.0509) & 0.2159(0.0270)  \\
          & True-LM  & 1.9547 & 2.8366(0.1082) & -0.0453(0.0267) & 1.9496  & 1.4553(0.0837) & -0.0504(0.0381) & 2.0175  & 0.7487(0.0475) & 0.0175(0.0261)  \\
          \midrule
$\beta_x$=1   & Adj-LM   & 1.1231 & 0.5178(0.0419) & 0.1231(0.0209)  & 1.0685  & 0.1564(0.0081) & 0.0685(0.0069)  & 1.0243  & 0.0684(0.0033) & 0.0243(0.0084)  \\
          & Naive-LM & 0.6693 & 0.2290(0.0087) & -0.3307(0.0097) & 0.6738  & 0.1576(0.0030) & -0.3262(0.0033) & 0.6677  & 0.1350(0.0039) & -0.3323(0.0054) \\
          & POI-SIMEX    & 0.9100 & 0.2442(0.0093) & -0.0900(0.0136) & 0.9160  & 0.1042(0.0035) & -0.0840(0.0047) & 0.9014  & 0.0569(0.0027) & -0.0986(0.0074) \\
          & True-LM  & 1.0121 & 0.1600(0.0060) & 0.0121(0.0149)  & 1.0119  & 0.0671(0.0039) & 0.0119(0.0047)  & 0.9984  & 0.0325(0.0018) & -0.0016(0.0048) \\
          \midrule
$\beta_z$=0.5 & Adj-LM   & 0.5058 & 0.0936(0.0045) & 0.0058(0.0079)  & 0.5019  & 0.0420(0.0024) & 0.0019(0.0065)  & 0.4997  & 0.0231(0.0016) & -0.0003(0.0045) \\
          & Naive-LM & 0.5077 & 0.0829(0.0030) & 0.0077(0.0069)  & 0.5032  & 0.0397(0.0022) & 0.0032(0.0058)  & 0.5007  & 0.0224(0.0015) & 0.0007(0.0041)  \\
          & POI-SIMEX    & 0.5073 & 0.0862(0.0032) & 0.0073(0.0072)  & 0.5026  & 0.0409(0.0024) & 0.0026(0.0062)  & 0.5000  & 0.0227(0.0016) & 0.0000(0.0044)  \\
          & True-LM  & 0.5065 & 0.0813(0.0028) & 0.0065(0.0072)  & 0.5017  & 0.0378(0.0019) & 0.0017(0.0062)  & 0.4995  & 0.0205(0.0015) & -0.0005(0.0044)\\
\bottomrule
\end{tabular*}
\begin{tablenotes}
\item[$^{\rm *}$] In all settings, mean squared error (MSE), Bias and Monte Carlo standard error (MCSE) are based on the 1000 simulated datasets.
\item[$^{\rm **}$] Adj-LM:Li \textit{et al.} \citeyearpar{Lietal2004} proposed bias correction method; Naive-LM:Without bias correction (i.e., use observed covariate);POI-SIMEX: Poisson SIMEX bias correction; True-LM: Use the true covariate from the simulation.
\end{tablenotes}
\bigskip\bigskip  
\caption{Simulation results for scenario 2:$X_i \sim Gamma(1,10)$, $W_i \lvert X_i \sim Poisson (X_i)$  and sample sizes 50, 100, and 200\label{tab2}}

\smallskip
\begin{tabular*}{\textwidth}{@{\extracolsep\fill}lllllllllll@{}}
\toprule
& &\multicolumn{3}{@{}l}{\textbf{N=50$^{\tnote{\bf *}}$}} & \multicolumn{3}{@{}l}{\textbf{N=100$^{\tnote{\bf *}}$}} & \multicolumn{3}{@{}l}{\textbf{N=200$^{\tnote{\bf *}}$}} \\\cmidrule{3-5}\cmidrule{6-8}\cmidrule{9-11}
\textbf{Parameter}& \textbf{Method$^{\tnote{\bf **}}$}  & \textbf{Estimate} & \textbf{MSE(MCSE)} & \textbf{Bias(MSSE)} & \textbf{Estimate} & \textbf{MSE(MCSE)} & \textbf{Bias(MCSE)} & \textbf{Estimate} & \textbf{MSE(MCSE)} & \textbf{Bias(MCSE)}\\
\midrule
$\beta_0$=2   & Adj-LM   & 1.9112 & 4.0535(0.2398) & -0.0888(0.0631) & 1.9159  & 2.1064(0.1545) & -0.0841(0.0491) & 1.9163  & 1.1559(0.0455) & -0.0837(0.0204) \\
          & Naive-LM & 2.9471 & 4.5793(0.2203) & 0.9471(0.0631)  & 2.8994  & 2.7660(0.1096) & 0.8994(0.0415)  & 2.8717  & 1.8024(0.0384) & 0.8717(0.0200)  \\
          & POI-SIMEX    & 2.0622 & 3.9916(0.2276) & 0.0622(0.0656)  & 1.9971  & 2.1163(0.1572) & -0.0029(0.0472) & 1.9794  & 1.1521(0.0429) & -0.0206(0.0193) \\
          & True-LM  & 1.9986 & 2.7964(0.1706) & -0.0014(0.0588) & 1.9897  & 1.4272(0.1096) & -0.0103(0.0355) & 1.9845  & 0.7636(0.0214) & -0.0155(0.0230) \\
\midrule
$\beta_x$=1   & Adj-LM   & 1.0121 & 0.0144(0.0007) & 0.0121(0.0040)  & 1.006   & 0.0062(0.0003) & 0.0060(0.0024)  & 1.0057  & 0.0035(0.0001) & 0.0057(0.0020)  \\
          & Naive-LM & 0.9061 & 0.0186(0.0008) & -0.0939(0.0034) & 0.9072  & 0.0133(0.0003) & -0.0928(0.0020) & 0.9100  & 0.0107(0.0004) & -0.0900(0.0018) \\
          & POI-SIMEX    & 0.9970  & 0.0131(0.0006) & -0.0030(0.0039) & 0.9973  & 0.0060(0.0002) & -0.0027(0.0023) & 0.9994  & 0.0034(0.0001) & -0.0006(0.0019) \\
          & True-LM  & 1.0000 & 0.0061(0.0004) & 0.0000(0.0015)  & 0.9999  & 0.0027(0.0001) & -0.0001(0.0014) & 1.0002  & 0.0014(0.0001) & 0.0002(0.0010)  \\
\midrule
$\beta_z$=0.5 & Adj-LM   & 0.5034 & 0.1126(0.0057) & 0.0034(0.0137)  & 0.5047  & 0.0572(0.0040) & 0.0047(0.0070)  & 0.5066  & 0.0272(0.0012) & 0.0066(0.0028)  \\
          & Naive-LM & 0.5037 & 0.1079(0.0060) & 0.0037(0.0130)  & 0.5037  & 0.0559(0.0038) & 0.0037(0.0065)  & 0.5058  & 0.0261(0.0013) & 0.0058(0.0032)  \\
          & POI-SIMEX    & 0.5020 & 0.1134(0.0057) & 0.0020(0.0139)  & 0.5054  & 0.0580(0.0042) & 0.0054(0.0070)  & 0.5064  & 0.0272(0.0012) & 0.0064(0.0030)  \\
          & True-LM  & 0.5068 & 0.0803(0.0040) & 0.0068(0.0100)  & 0.5041  & 0.0401(0.0025) & 0.0041(0.0057)  & 0.5031  & 0.0191(0.0009) & 0.0031(0.0029) \\
\bottomrule
\end{tabular*}
\begin{tablenotes}
\item[$^{\rm *}$] In all settings, mean squared error (MSE), Bias and Monte Carlo standard error (MCSE) are based on the 1000 simulated datasets.
\item[$^{\rm **}$] Adj-LM:Li \textit{et al.} \citeyearpar{Lietal2004} proposed bias correction method; Naive-LM:Without bias correction (i.e., use observed covariate);POI-SIMEX: Poisson SIMEX bias correction; True-LM: use the true covariate from the simulation.
\end{tablenotes}
\end{sidewaystable}

\begin{sidewaystable}
\def\d{\hphantom{0}}
\caption{Simulation results for scenario 3: $X_i \sim Gamma(2,Z), Z \sim Uniform(0.5,9), W_i \lvert X_i \sim Poisson (X_i)$ and sample sizes 50, 100, and 200\label{tab3}}
\begin{tabular*}{\textwidth}{@{\extracolsep\fill}lllllllllll@{}}
\toprule
& &\multicolumn{3}{@{}l}{\textbf{N=50$^{\tnote{\bf *}}$}} & \multicolumn{3}{@{}l}{\textbf{N=100$^{\tnote{\bf *}}$}} & \multicolumn{3}{@{}l}{\textbf{N=200$^{\tnote{\bf *}}$}} \\\cmidrule{3-5}\cmidrule{6-8}\cmidrule{9-11}
\textbf{Parameter}& \textbf{Method$^{\tnote{\bf **}}$}  & \textbf{Estimate} & \textbf{MSE(MCSE)} & \textbf{Bias(MCSE)} & \textbf{Estimate} & \textbf{MSE(MCSE)} & \textbf{Bias(MCSE)} & \textbf{Estimate} & \textbf{MSE(MCSE)} & \textbf{Bias(MCSE)}\\
\midrule
$\beta_0$=2   & Adj-LM   & 1.9721 & 2.9498(0.0632) & -0.0279(0.0670) & 1.9717  & 1.4085(0.0506) & -0.0283(0.0385) & 2.0037  & 0.7688(0.0445) & 0.0037(0.0284)  \\
          & Naive-LM & 1.9950 & 2.7955(0.0723) & -0.0050(0.0700) & 1.9845  & 1.3651(0.0488) & -0.0155(0.0374) & 2.0099  & 0.7451(0.0428) & 0.0099(0.0274)  \\
          & POI-SIMEX    & 1.9753 & 2.9162(0.0593) & -0.0247(0.0699) & 1.9726  & 1.3946(0.0541) & -0.0274(0.0375) & 2.0100  & 0.7647(0.0438) & 0.0100(0.0291)  \\
          & True-LM  & 1.9645 & 2.4152(0.0879) & -0.0355(0.0731) & 1.9879  & 1.1474(0.0375) & -0.0121(0.0247) & 1.9956  & 0.6462(0.0303) & -0.0044(0.0259) \\
\midrule
$\beta_x$=1   & Adj-LM   & 1.0298 & 0.0285(0.0016) & 0.0298(0.0039)  & 1.0121  & 0.0117(0.0004) & 0.0121(0.0037)  & 1.0049  & 0.0055(0.0002) & 0.0049(0.0021)  \\
          & Naive-LM & 0.8550 & 0.0364(0.0011) & -0.1450(0.0032) & 0.8580  & 0.0275(0.0008) & -0.1420(0.0027) & 0.8617  & 0.0227(0.0005) & -0.1383(0.0018) \\
          & POI-SIMEX    & 0.9943 & 0.0232(0.0011) & -0.0057(0.0038) & 0.9906  & 0.0106(0.0003) & -0.0094(0.0034) & 0.9892  & 0.0052(0.0002) & -0.0108(0.0021) \\
          & True-LM  & 1.0009 & 0.0106(0.0007) & 0.0009(0.0033)  & 1.0012  & 0.0046(0.0002) & 0.0012(0.0016)  & 0.9981  & 0.0021(0.0001) & -0.0019(0.0017) \\
 \midrule
$\beta_z$=0.5 & Adj-LM   & 0.4458 & 0.1888(0.0099) & -0.0542(0.0120) & 0.4748  & 0.0820(0.0034) & -0.0252(0.0130) & 0.4916  & 0.0427(0.0020) & -0.0084(0.0066) \\
          & Naive-LM & 0.7851 & 0.2243(0.0132) & 0.2851(0.0128)  & 0.7774  & 0.1449(0.0067) & 0.2774(0.0120)  & 0.7757  & 0.1114(0.0023) & 0.2757(0.0059)  \\
          & POI-SIMEX    & 0.5139 & 0.1718(0.0103) & 0.0139(0.0122)  & 0.5169  & 0.0786(0.0034) & 0.0169(0.0126)  & 0.5219  & 0.0417(0.0016) & 0.0219(0.0067)  \\
          & True-LM  & 0.5050 & 0.1187(0.0069) & 0.0050(0.0092)  & 0.4940  & 0.0540(0.0029) & -0.0060(0.0083) & 0.5062  & 0.0271(0.0010) & 0.0062(0.0051) \\
\bottomrule
\end{tabular*}
\begin{tablenotes}
\item[$^{\rm *}$] In all settings, mean squared error (MSE), Bias and Monte Carlo standard error (MCSE) are based on the 1000 simulated datasets.
\item[$^{\rm **}$] Adj-LM:Li \textit{et al.} \citeyearpar{Lietal2004} proposed bias correction method; Naive-LM:Without bias correction (i.e., use observed covariate);POI-SIMEX: Poisson SIMEX bias correction; True-LM: use the true covariate from the simulation.
\end{tablenotes}
\bigskip\bigskip  
\caption{Simulation results under different ratios of the variance of true covariate over the observed covariate \label{tab4}}

\smallskip
\begin{tabular*}{\textwidth}{@{\extracolsep\fill}lllllllllll@{}}
\toprule
& &\multicolumn{3}{@{}l}{\textbf{Gamma(0.1,9)$^{\tnote{\bf *}}$}} & \multicolumn{3}{@{}l}{\textbf{Gamma(2/3,3)$^{\tnote{\bf *}}$}} & \multicolumn{3}{@{}l}{\textbf{Gamma(2,1)$^{\tnote{\bf *}}$}} \\\cmidrule{3-5}\cmidrule{6-8}\cmidrule{9-11}
\textbf{Parameter}& \textbf{Method$^{\tnote{\bf **}}$} & \textbf{Estimate} & \textbf{MSE(MCSE)} & \textbf{Bias(MCSE)} & \textbf{Estimate} & \textbf{MSE(MCSE)} & \textbf{Bias(MCSE)} & \textbf{Estimate} & \textbf{MSE(MCSE)} & \textbf{Bias(MCSE)}\\
\midrule
$\beta_0=2$   & Adj-LM   & 1.9538       & 1.2186( 0.0659) & -0.0462(0.0441) & 1.9776       & 1.5991( 0.0389) & -0.0224(0.0360) & 1.7675     & 3.5009( 0.2143) & -0.2325(0.0683) \\
          & Naive-LM & 2.0742       & 1.1820( 0.0623) & 0.0742(0.0430)  & 2.5503       & 1.6960( 0.0555) & 0.5503(0.0318)  & 2.9984     & 2.5109( 0.1145) & 0.9984(0.0507)  \\
          & POI-SIMEX    & 1.9730       & 1.2041( 0.0635) & -0.0270(0.0435) & 2.1359       & 1.5322( 0.0348) & 0.1359(0.0347)  & 2.5023     & 2.2108( 0.0935) & 0.5023(0.0548)  \\
          & True-LM  & 1.9794       & 1.1468( 0.0673) & -0.0206(0.0426) & 2.0641       & 1.3416( 0.0503) & 0.0641(0.0360)  & 1.9898     & 1.7354( 0.0748) & -0.0102(0.0556) \\
    \midrule
$\beta_x=1$  & Adj-LM   & 1.0513       & 0.1184( 0.0072) & 0.0513(0.0128)  & 1.0481       & 0.0991( 0.0048) & 0.0481(0.0093)  & 1.0951     & 0.5245( 0.0268) & 0.0951(0.0185)  \\
          & Naive-LM & 0.8969       & 0.0784( 0.0034) & -0.1031(0.0097) & 0.7561       & 0.1027( 0.0025) & -0.2439(0.0054) & 0.4861     & 0.3332( 0.0070) & -0.5139(0.0064) \\
          & POI-SIMEX    & 1.0257       & 0.1006( 0.0046) & 0.0257(0.0118)  & 0.9656       & 0.0757( 0.0031) & -0.0344(0.0077) & 0.7287     & 0.2341( 0.0091) & -0.2713(0.0100) \\
          & True-LM  & 0.9963       & 0.0549( 0.0023) & -0.0037(0.0073) & 1.0018       & 0.0462( 0.0014) & 0.0018(0.0064)  & 0.9891     & 0.1369( 0.0060) & -0.0109(0.0097) \\
    \midrule
$\beta_z=0.5$ & Adj-LM   & 0.5053       & 0.0400( 0.0022) & 0.0053(0.0075)  & 0.4872       & 0.0452( 0.0016) & -0.0128(0.0055) & 0.5115     & 0.0463( 0.0027) & 0.0115(0.0088)  \\
          & Naive-LM & 0.5051       & 0.0393( 0.0022) & 0.0051(0.0075)  & 0.4868       & 0.0434( 0.0017) & -0.0132(0.0052) & 0.5065     & 0.0395( 0.0021) & 0.0065(0.0092)  \\
          & POI-SIMEX    & 0.5047       & 0.0399( 0.0021) & 0.0047(0.0075)  & 0.4869       & 0.0445( 0.0017) & -0.0131(0.0053) & 0.5088     & 0.0411( 0.0022) & 0.0088(0.0090)  \\
          & True-LM  & 0.5065       & 0.0378( 0.0023) & 0.0065(0.0072)  & 0.4851       & 0.0411( 0.0014) & -0.0149(0.0052) & 0.5058     & 0.0390( 0.0021) & 0.0058(0.0090) \\
\bottomrule
\end{tabular*}
\begin{tablenotes}
\item[$^{\rm *}$] In all settings, N=100, mean squared error (MSE), Bias,  and Monte Carlo standard error (MCSE) are based on the 1000 simulated datasets using batch mean method with 10 batches. Ratios (left to right): 0.9, 0.75, 0.5, defined as $\frac{\sigma_X^2}{\sigma_W^2}$.
\item[$^{\rm **}$] Adj-LM:Li \textit{et al.} \citeyearpar{Lietal2004} proposed bias correction method; Naive-LM: Without bias correction (i.e., use observed covariate); POI-SIMEX: Poisson SIMEX bias correction; True-LM: use the true covariate from the simulation.
\end{tablenotes}

\end{sidewaystable}

We also perform simulation studies to investigate the performance under different ratios, defined as the ratio of the variance of true covariate ($\sigma_X^2$) and the variance of the observed covariate ($\sigma_W^2$). The data are simulated using the settings described above,  except that the Gamma shape and scale parameters are chosen such that the ratio, defined as $\frac{\sigma_X^2}{\sigma_W^2}$ [=$ab^2/(ab+ab^2)]$ in the conditional Poisson error model given the Gamma distributed true covariate, are (1) $r = 0.9$, for $a = 0.1, b = 9$; (2) $r = 0.75$, for $a = 2/3, b = 3$; (3) $r = 0.5$, for $a = 2, b = 1$. The sample size $N$ is $100$.

For each setting, the bias and MSE, and their Monte Carlo standard errors, are reported for 1000 data sets. The results corresponding to the covariate with measurement error are summarized in Table~\ref{tab4}.

In scenario 1, when the variance of the true covariate is relatively large compared to its mean, the Naive method has the smallest MSE but a significant bias. If we consider both MSE and bias, POI-SIMEX exhibits the second smallest MSE and the smallest bias.  In scenario 2, POI-SIMEX outperforms the Adj-LM method with the smallest MSE and bias. In scenario 3, the Adj-LM method has the largest MSE, whereas POI-SIMEX has the smallest MSE. 

In the Appendix we report on four additional simulation studies that investigate performance under model misspecification, where the observed covariate conditional on the true covariate is drawn from either a binomial distribution (two studies) or a negative binomial distribution (two studies) and the conditional Poisson approach is applied. All four studies represent very strong deviations from the model assumptions, with the specific simulation settings and results reported in Appendix Tables \ref{atab1} - \ref{atab4}. As might be expected, the performance of both approaches based on the conditional Poisson assumption exhibit performance degradation under strong forms of model misspecification. Even in these cases, however, we observe that the degradation of POI-SIMEX is generally less severe than the degradation of Adj-LM, and thus it exhibits improved performance under misspecification as well. 

\subsection{Simulation with log-normal censored survival data}\label {subsim_lognorm}
We evaluate the performance of POI-SIMEX with data generated using an accelerated failure time survival model with a conditionally Poisson-distributed error-prone covariate. We simulate log-normal survival times with 20 percent random censoring. The Poisson rate $X_i$s are generated independently from a Gamma distribution with shape parameter, $a=1$, and scale parameter, $b=2$. The scale parameter ($\sigma$) for the log-normal distribution of the error is set to 2. The covariate without measurement error, $Z$, is drawn from a uniform distribution between 0.5 and 9. The intercept and coefficients associated with log survival time are $\beta_0=2$, $\beta_x=1$, and $\beta_z=0.5$. The sample size N is 50, 100, or 200. In all instances, the bias, mean squared error (MSE), and their Monte Carlo standard errors, are reported from 1000 datasets generated. The results for all parameters are summarized in Table ~\ref{tab5}. We compare POI-SIMEX only with the Naive method as Adj-LM can not be applied in settings with censoring. However, POI-SIMEX is easily adapted to this and other regression settings with Poisson surrogates. The POI-SIMEX method outperforms the Naive method with a smaller bias and mean squared error (MSE) in all scenarios. For instance, when $N=50$, the bias correction is significant for the intercept term ($\beta_0$), the coefficient of the error-prone covariate  ($\beta_X$), and the scale parameter ($\sigma$). With $\beta_0=2$, the naive method has an average estimate of $3.105$, while POI-SIMEX has an average estimate of $2.596$. For $\beta_X=1$, the naive method has an average estimate of $0.673$, whereas POI-SIMEX has an average estimate of $0.914$. For $\sigma=2$, the Naive method gives an average estimate of $2.364$, and POI-SIMEX gives an average estimate of $2.160$. Similar patterns are observed for $N=100$ and $N=200$. Both methods closely estimate the $\beta_Z$, as expected due to the assumed independence between covariates. 
\begin{sidewaystable}
\def\d{\hphantom{0}}
\caption{Simulation for survival data (20\% random right censoring) from log-normal with scale parameter ($\sigma=2$) and sample sizes 50, 100, and 200\label{tab5}}
\begin{tabular*}{\textwidth}{@{\extracolsep\fill}lllllllllll@{}}
\toprule
& &\multicolumn{3}{@{}l}{\textbf{N=50$^{\tnote{\bf *}}$}} & \multicolumn{3}{@{}l}{\textbf{N=100$^{\tnote{\bf *}}$}} & \multicolumn{3}{@{}l}{\textbf{N=200$^{\tnote{\bf *}}$}} \\\cmidrule{3-5}\cmidrule{6-8}\cmidrule{9-11}
\textbf{Parameter}& \textbf{Method$^{\tnote{\bf **}}$}  & \textbf{Estimate} & \textbf{MSE(MCSE)} & \textbf{Bias(MCSE)} & \textbf{Estimate} & \textbf{MSE(MCSE)} & \textbf{Bias(MCSE)} & \textbf{Estimate} & \textbf{MSE(MCSE)} & \textbf{Bias(MCSE)}\\
\midrule
$\beta_0=2$   & Naive-AFT & 3.1055 & 1.8954(0.0609) & 1.1055(0.0233)  & 3.1394  & 1.6334(0.0470) & 1.1394(0.0212)  & 3.1504  & 1.5199(0.0273) & 1.1504(0.0116)  \\
          & SIMEX-AFT & 2.5960 & 1.1946(0.0497) & 0.5960(0.0244)  & 2.6288  & 0.8033(0.0325) & 0.6288(0.0254)  & 2.6446  & 0.6464(0.0182) & 0.6446(0.0130)  \\
          & True-AFT  & 2.3963 & 0.7038(0.0340) & 0.3963(0.0167)  & 2.4136  & 0.4399(0.0196) & 0.4136(0.0195)  & 2.4174  & 0.3276(0.0108) & 0.4174(0.0106)  \\
\midrule
$\beta_x=1$   & Naive-AFT & 0.6731 & 0.1440(0.0033) & -0.3269(0.0058) & 0.6747  & 0.1230(0.0032) & -0.3253(0.0051) & 0.6712  & 0.1160(0.0010) & -0.3288(0.0012) \\
          & SIMEX-AFT & 0.9138 & 0.0845(0.0028) & -0.0862(0.0089) & 0.9109  & 0.0426(0.0023) & -0.0891(0.0071) & 0.9040   & 0.0251(0.0009) & -0.0960(0.0020) \\
          & True-AFT  & 1.0059 & 0.0312(0.0016) & 0.0059(0.0068)  & 1.0019  & 0.0145(0.0009) & 0.0019(0.0035)  & 1.0026  & 0.0067(0.0004) & 0.0026(0.0019)  \\
          \midrule
$\beta_z=0.5$ & Naive-AFT & 0.4993 & 0.0217(0.0010) & -0.0007(0.0041) & 0.4997  & 0.0097(0.0006) & -0.0003(0.0039) & 0.4975  & 0.0055(0.0002) & -0.0025(0.0019) \\
          & SIMEX-AFT & 0.4991 & 0.0224(0.0010) & -0.0009(0.0036) & 0.5003  & 0.0099(0.0006) & 0.0003(0.0040)  & 0.4978  & 0.0057(0.0002) & -0.0022(0.0018) \\
          & True-AFT  & 0.4977 & 0.0158(0.0008) & -0.0023(0.0026) & 0.5011  & 0.0073(0.0004) & 0.0011(0.0037)  & 0.4994  & 0.0042(0.0002) & -0.0006(0.0015) \\
          \midrule
$\sigma=2 $  & Naive-AFT & 2.3642 & 0.2118(0.0100) & 0.3642(0.0095)  & 2.4232  & 0.2189(0.0033) & 0.4232(0.0043)  & 2.4339  & 0.2089(0.0027) & 0.4339(0.0034)  \\
          & SIMEX-AFT & 2.1599 & 0.1263(0.0075) & 0.1599(0.0101)  & 2.2163  & 0.0955(0.0028) & 0.2163(0.0061)  & 2.2258  & 0.0761(0.0015) & 0.2258(0.0033)  \\
          & True-AFT  & 2.0499 & 0.0571(0.0024) & 0.0499(0.0098)  & 2.1019  & 0.0382(0.0015) & 0.1019(0.0042)  & 2.109   & 0.0259(0.0008) & 0.1090(0.0034) \\
\bottomrule
\end{tabular*}
\begin{tablenotes}
\item[$^{\rm *}$] In all settings, true $X \sim$ Gamma(1,2), mean squared error (MSE), Bias, and Monte Carlo standard error (MCSE) are based on the 1000 simulated datasets.
\item[$^{\rm **}$] Naive-AFT:log-normal AFT model without bias correction (i.e., use observed covariate); SIMEX-AFT: log-normal AFT model using POI-SIMEX for bias correction; True-AFT: log-normal use the true covariate from the simulation.
\end{tablenotes}
\end{sidewaystable}

\section{Application to the COEUR HGSC cohort} \label{sec5}

\begin{figure*}
\centerline{\includegraphics[scale=0.85]{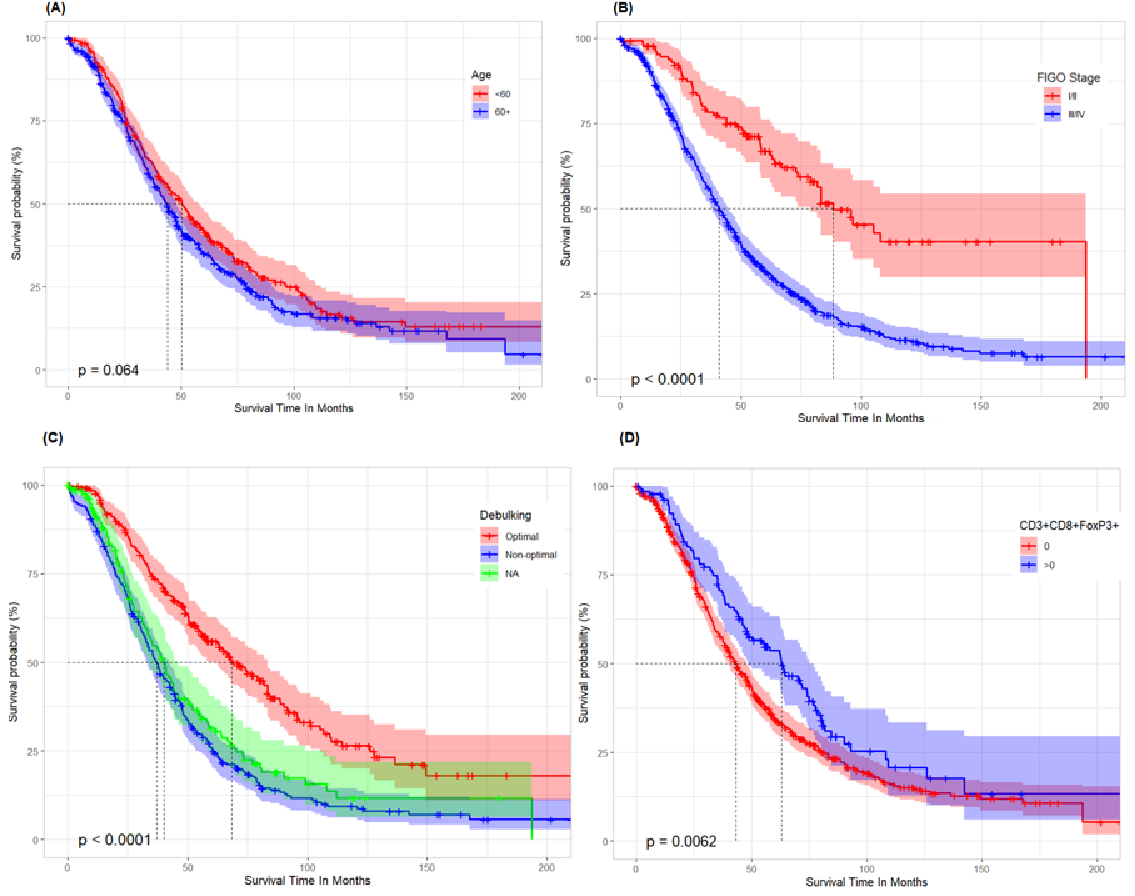}}
\caption{Kaplan-Meier survival curves with 95\% confidence interval band, log-rank test p-value, and median survival time (dash line) by strata\label{fig2}}
\end{figure*}

Despite being ten times less common than breast cancer, epithelial ovarian cancer (EOC) ranks as the fifth-leading cause of cancer-related deaths among women in Canada \citep{Arora2018}. Over the past decade, advancements in therapeutic targets and drug developments have notably increased survival rates. However, despite these improvements, patients' survival rates and the overall quality of life remain suboptimal. High-grade serous carcinoma (HGSC) is a subtype of EOC. Epithelial ovarian cancer is a broad category of cancers that arise from the epithelial cells covering the surface of the ovaries or, as current research suggests, the fallopian tube \citep{Labidi2017}. It includes several subtypes, with high-grade serous carcinoma being one of the most common and aggressive forms. Thus, there is a pressing need to find biomarkers that are relevant to survival. COEUR comprises 12 ovarian cancer biobanks across Canada. This repository includes a central retrospective cohort of over 2,000 patient tissue samples with associated clinical data, of which 1,246 are HGSC. Our study cohort of 716 is a subset of the HGSC samples, that only includes patients with linked TMA data, complete clinical information, and demographic details such as treatment, survival, or follow-up time. For the purpose of this analysis we assume that the missingness is ignorable. We apply multivariable linear regression, assuming the log survival time is normally distributed and correct the bias in coefficient estimation using POI-SIMEX. Table (\ref{tab6}) displays summary statistics of subject characteristics stratified by death status, and Kaplan-Meier survival curves in Figure (\ref{fig2}) show that optimal debulking, early FIGO stage (I or II), and presenting CD3+CD8+FoxP3+ $\sim$E ($\sim$E refers to the cells measured in epithelial regions) are positively associated with survival. 

The CD3+CD8+FoxP3+ $\sim$E triple-positive biomarker, derived from TMA data, highlights the critical role of immune cell infiltration, such as tumor-infiltrating lymphocytes (TILs), which are key prognostic indicators for survival across various cancers. Regulatory T cells (Tregs), characterized by FoxP3 expression, are primarily recognized for their immunosuppressive function, maintaining immune homeostasis by downregulating immune responses and preventing autoimmunity. However, in the tumor microenvironment, Tregs' suppressive activity can hinder anti-tumor immunity, often correlating with poor prognosis \citep{Curiel2004,Curiel2008,Elkoshi2022}. In some cancer subtypes, including ovarian cancer, immune infiltration has been linked to overall prognosis. Specifically, CD8+FoxP3+ T cells have been suggested to form a unique regulatory population with suppressive effects on infiltrating B cells \citep{Preston2013,Liston2022,Warfvinge2017}. In advanced-stage ovarian cancer, the presence of CD8+ cells, FoxP3+ Tregs, or a high CD8/FoxP3 ratio in tumor tissue has been linked to improved disease-specific survival \citep{Leffers2009}. Similarly, in high-grade serous tumors from optimally debulked patients, disease-specific survival was positively correlated with CD8 and FoxP3 markers \citep{Milne2009}. Nevertheless, the specific impact of these triple-positive cells on HGSC patient survival remains unclear and and we hope that the results from applying POI-SIMEX will provide further insight. Notably, the estimated coefficient of the biomarker obtained from POI-SIMEX is larger than the estimate obtained from the naive model with a slightly increased standard error relative to the naive approach (Table \ref{tab7}). We have added the standard error from the bootstrap method to the POI-SIMEX estimator, which is comparable with the standard error obtained using SIMEX and its standard variance estimation. As shown in Table \ref{tab7}, for each unit increase in the density (per $mm^2$) of CD3+CD8+FoxP3+ $\sim$E, the estimated increase in log survival time is 0.0148 (SE:0.0053), given that the other covariates are held fixed. Compared to the naive method, this represents a relative increase in the estimated regression coefficient of approximately 25 percent. There is clearly a non-trivial correction to the estimate associated with the biomarker. This highlights the importance of properly accounting for measurement error when evaluating the prognostic value of immune biomarkers.

\begin{center}
\begin{table*}[!h]%
\caption{Summary statistics of patient characteristics stratified by death event status for COEUR HGSOC cohort\label{tab6}}
 \begin{tabular*}{\textwidth}{@{\extracolsep\fill}llllll@{}}
\toprule    
& &Alive & Dead & Total &\\
&CD3+CD8+FoxP3+ $\sim$E$^{\tnote{*}}$ & & & &\\  
&~ ~ 0  & 159   & 424   & 583 &  \\
&~ ~ >0   & 50    & 83    & 133 &  \\
&         &       &       &     &  \\
&Age (60+)    & & & &\\
&~ ~ Yes & 119   & 297  & 416  & \\
&~ ~ No  & 90    & 210   & 300  & \\
 &       &       &       &    &   \\
&Non-optimal debulking     & & & & \\
&~ ~ Yes       & 42    & 244   & 286  & \\
&~ ~ No        & 115   & 151   & 266  & \\
&~ ~ NA        & 52    & 112   & 164 & \\
 &         &       &       &     &  \\
&FIGO stage (III/IV)   & & & &\\  
&~ ~ Yes  & 134   & 450   & 584 &  \\
&~ ~ No   & 75    & 57    & 132 &  \\
&         &       &       &     &  \\
\bottomrule
\end{tabular*}
\begin{tablenotes}
\item[$^{\rm *}$] Indication of the expression of Tregs CD3+CD8+FoxP3+ in the epithelial area
\end{tablenotes}
\end{table*}
\end{center}

\begin{center}
\begin{table*}[!ht]%
\caption{Log-normal accelerated survival time regression with Poisson distributed biomarker CD3+CD8+FoxP3+ $\sim$E \label{tab7}}
\begin{tabular*}{\textwidth}{@{\extracolsep\fill}lllllllll@{}}
\toprule
&\multicolumn{3}{@{}l}{\textbf{Naive$^{\tnote{\bf a}}$}} & \multicolumn{5}{@{}l} {\textbf{POI-SIMEX$^{\tnote{\bf b}}$}}
\\\cmidrule{2-4}\cmidrule{5-9}
\textbf{Parameters} & \textbf{Estimate} & \textbf{SE} & \textbf{p-value} & \textbf{Estimate} & \textbf{SE} & \textbf{SE(Bootstrap)} & \textbf{p-value} & \textbf{$\lvert$ rel.pct chg.$\lvert$$^{\tnote{\bf c}}$}\\
\midrule
Intercept                 & 4.8238   & 0.1298 & \textless{}0.001 & 4.8191   & 0.1297 & 0.1248         & \textless{}0.001  & 0.0974\\
CD3+CD8+FoxP3+ $\sim$E$^{\tnote{*}}$ & 0.0118   & 0.0041 & ~0.004            & 0.0148   & 0.0053 & 0.0070   & ~0.005     & 25.4237       \\
Age 60+      & -0.2131  & 0.0844 & ~0.012            & -0.2136  & 0.0845 & 0.0760       & ~0.011         & 0.2346   \\
Non-optimal debulking (Yes)          & -0.5967  & 0.1072 & \textless{}0.001 & -0.5963  & 0.1071 & 0.0891         & \textless{}0.001 & 0.0670\\
Non-optimal debulking (NA) & -0.4722  & 0.1071 & \textless{}0.001 & -0.4735  & 0.1072 & 0.1031         & \textless{}0.001 &0.2753 \\
FIGO stage (III/IV) & -0.6628  & 0.1312 & \textless{}0.001 & -0.6653  & 0.1311 & 0.1199         & \textless{}0.001 &0.3772\\
\bottomrule
\end{tabular*}
\begin{tablenotes}
\item[$^{\rm *}$] Indication of the expression of Tregs CD8+FoxP3+ in the epithelial area
\item[$^{\rm a}$] Estimated scale parameter of lognormal:1.3126.
\item[$^{\rm b}$] Estimated scale parameter of lognormal:1.0600.
\item [$^{\rm c}$] The absolute relative percent change is computed by $100*\lvert$ (Estimate$_{POI-SIMEX} - $Estimate$_{Naive}$)/Estimate$_{Naive}\lvert$
\end{tablenotes}
\end{table*}
\end{center}

\section{Conclusion}\label{sec6}
 Motivated by the importance of adjusting for measurement error in TMA-based studies of cancer, we develop POI-SIMEX, an adjustment approach for handling conditionally Poisson-distributed surrogates, and applied it to linear regression models and parametric accelerated failure time survival models. For cases with unknown measurement error variance, our proposed method only requires one replicate to obtain a strongly consistent estimator of the measurement error variance and of the regression coefficient associated with the TMA-based covariate. Our simulation studies show that the proposed method demonstrates improved finite sample performance over existing methodology within the setting of linear regression and also has broad applicability for the analysis of TMAs beyond linear regression. This is demonstrated in simulations and an application involving censored survival data with conditionally Poisson surrogates. The POI-SIMEX approach is also appealing as it does not require specification of the true covariate process. While performance degradation is observed under model misspecification, the degree of degradation is less severe than that of competing methodology. 

Biologically, the finding that CD3+CD8+FOXP3+ T cells may be associated with increased overall survival is intriguing. Though we note that the effect size is relatively small and the association is only marginally significant with $p=0.005$ \citep{Benjamin2018}. One might speculate that these FOXP3+ T cells mediate direct killing of tumor cells, as FOXP3 is a marker of T cell activation. Alternatively, these cells could play an immunoregulatory role by reducing inflammatory signals such as TNF-a, a critical tumor factor, thereby indirectly blocking tumor growth and metastasis \citep{Ostrand2010}. It may also be possible that these cells limit the activity of myeloid-derived suppressor cells \citep{Gabrilovic2009}. Further functional studies are  needed clarify these potential scenarios \citep{Liu2015}.

This paper initiates the evaluation of POI-SIMEX through simulation experiments and its application within a parametric log-normal survival analysis framework. Extension to more general settings, including semiparametric survival models such as the Cox proportional hazards model, represents an avenue for future research. Through detailed finite-sample assessments and comparative analyses using real-world data, the proposed methodology offers a robust framework for enhancing the statistical analysis of tissue microarray (TMA)-based investigations in biomarker discovery, immunological profiling, and cancer survival studies. Companion R software implementing POI-SIMEX is publicly available via GitHub (\url{https://github.com/Aijunyan/POI-SIMEX}).

\begin{acknowledgement}
We thank the patients who contributed tumor samples and clinical data to the COEUR cohort.
Nathoo acknowledges funding from CIHR (PJT$-186124$) and the Terry Fox Research Institute (TFRI$1136-03$).
Nathoo and Lesperance acknowledge funding from NSERC (RGPIN$-04044-2020$, RGPIN$-07079-2020$) through the Discovery Grants Program.  Yang acknowledges funding from the University of Victoria and the Pacific Leaders Scholarship Program. Nelson acknowledges the BC Cancer Foundation, CIHR (PJT$-427647$),  and Terry Fox Research Institute (TFRI$1060$)for funding. Lum acknowledges funding from CIHR (PJT$-192015$), Terry Fox Research Institute, and Lotte \& John Hecht Memorial Foundation (TFRI$1125$). 

\end{acknowledgement}
\vspace*{1pc}

\noindent {\bf{Conflict of Interest}}

\noindent {\it{The authors have declared no conflict of interest.}}

\section*{Appendix --Simulation on misspecified models}\label{secA1}

\setcounter{table}{0}

To evaluate the model performance under misspecification, we conduct two simulation studies to evaluate the model's performance when the conditional Poisson assumption does not hold for covariates with measurement error. The same parameters from Section \ref{subsim} are used. However, we assume one scenario where observations are drawn from a binomial distribution and another from a negative binomial distribution with the mean set to the true covariate.

\begin{sidewaystable}
\def\d{\hphantom{0}}
\caption{Appendix: Simulation results for misspecification:  $X_i \sim Gamma(a,b), W_i \lvert X_i \sim Binormial (40,P_i)$ where $P_i=X_i/40$, and sample size 100 \label{atab1}}
\smallskip
\begin{tabular*}{\textwidth}{@{\extracolsep\fill}lllllllllll@{}}
\toprule
& &\multicolumn{3}{@{}l}{\textbf{Gamma(1,2)$^{\tnote{\bf *}}$}} & \multicolumn{3}{@{}l}{\textbf{Gamma(1,10)$^{\tnote{\bf *}}$}} & \multicolumn{3}{@{}l}{\textbf{Gamma(2,Z), $Z \sim Uniform(0.5,9)$$^{\tnote{\bf *}}$}} \\\cmidrule{3-5}\cmidrule{6-8}\cmidrule{9-11}
\textbf{Parameter}& \textbf{Method$^{\tnote{\bf **}}$}  & \textbf{Estimate} & \textbf{MSE(MCSE)} & \textbf{Bias(MCSE)} & \textbf{Estimate} & \textbf{MSE(MCSE)} & \textbf{Bias(MCSE)} & \textbf{Estimate} & \textbf{MSE(MCSE)} & \textbf{Bias(MCSE)}\\
\midrule
$\beta_0=2$   & Adj-LM   & 1.7553     & 1.9951 (0.099)  & -0.2447 (0.0556) & 1.0037      & 3.3613 (0.1679) & -0.9963 (0.0477) & 1.8882   & 1.3779 (0.0719) & -0.1118 (0.0229) \\
          & Naive-LM & 2.5795     & 1.7739 (0.0785) & 0.5795 (0.0476)  & 2.2435      & 2.1818 (0.101)  & 0.2435 (0.0446)  & 1.9232  & 1.2978 (0.0628) & -0.0768 (0.0267) \\
          & POI-SIMEX    & 2.0796     & 1.6822 (0.0673) & 0.0796 (0.0503)  & 1.1186      & 3.1261 (0.1625) & -0.8814 (0.0481) & 1.8895  & 1.3757 (0.075)  & -0.1105 (0.0238) \\
          & True-LM  & 1.9708     & 1.4294 (0.0468) & -0.0292 (0.0456) & 2.0488      & 1.5041 (0.0842) & 0.0488 (0.0336)  & 2.0478         & 1.0983 (0.0552) & 0.0478 (0.0227)  \\
\midrule
$\beta_x=1$   & Adj-LM   & 1.1147     & 0.1708 (0.0105) & 0.1147 (0.0158)  & 1.1251      & 0.0244 (0.0009) & 0.1251 (0.0031)  & 1.1603                           & 0.0408 (0.0014) & 0.1603 (0.0034)  \\
          & Naive-LM & 0.6937     & 0.1438 (0.0051) & -0.3063 (0.0084) & 0.9980       & 0.0074 (0.0004) & -0.002 (0.0028)  & 0.9556                           & 0.0122 (0.0002) & -0.0444 (0.0028) \\
          & POI-SIMEX    & 0.9469     & 0.1002 (0.0038) & -0.0531 (0.0122) & 1.1128      & 0.0213 (0.0008) & 0.1128 (0.003)   & 1.1277                           & 0.0303 (0.0011) & 0.1277 (0.0033)  \\
          & True-LM  & 1.0010      & 0.0673 (0.0026) & 0.0010 (0.0103)   & 0.9993      & 0.0027 (0.0002) & -0.0007 (0.0019) & 1.0019                           & 0.0045 (0.0002) & 0.0019 (0.0012)  \\
\midrule
$\beta_z=0.5$ & Adj-LM   & 0.5038     & 0.0434 (0.0017) & 0.0038 (0.0079)  & 0.4914      & 0.0601 (0.0025) & -0.0086 (0.006)  & 0.2304                           & 0.1643 (0.0061) & -0.2696 (0.0068) \\
          & Naive-LM & 0.5048     & 0.0407 (0.0016) & 0.0048 (0.0079)  & 0.4909      & 0.0582 (0.0023) & -0.0091 (0.0054) & 0.6258                           & 0.0824 (0.0038) & 0.1258 (0.0063)  \\
          & POI-SIMEX    & 0.5044     & 0.0419 (0.0017) & 0.0044 (0.0079)  & 0.4925      & 0.0596 (0.0025) & -0.0075 (0.0058) & 0.2935                           & 0.1287 (0.0054) & -0.2065 (0.0064) \\
          & True-LM  & 0.5038     & 0.0384 (0.0013) & 0.0038 (0.0075)  & 0.4943      & 0.0428 (0.0022) & -0.0057 (0.0031) & 0.4857                           & 0.0525 (0.0029) & -0.0143 (0.0058)\\
\bottomrule
\end{tabular*}
\begin{tablenotes}
\item[$^{\rm *}$] In all settings, mean squared error (MSE), Bias and Monte Carlo standard error (MCSE) are based on the 1000 simulated datasets.
\item[$^{\rm **}$] Adj-LM:Li \textit{et al.}\citeyearpar{Lietal2004} proposed bias correction method; Naive-LM:Without bias correction (i.e., use observed covariate);POI-SIMEX: Poisson SIMEX bias correction; True-LM: use the true covariate from the simulation.
\end{tablenotes}
\bigskip\bigskip  
\caption{Appendix: Simulation results for misspecification:  $X_i \sim Gamma(a,b), W_i \lvert X_i \sim Binomial (60,P_i)$ where $P_i=X_i/60$, and sample size 100 \label{atab2}}
\smallskip
\begin{tabular*}{\textwidth}{@{\extracolsep\fill}lllllllllll@{}}
\toprule
& &\multicolumn{3}{@{}l}{\textbf{Gamma(1,2)$^{\tnote{\bf *}}$}} & \multicolumn{3}{@{}l}{\textbf{Gamma(10,1)$^{\tnote{\bf *}}$}} & \multicolumn{3}{@{}l}{\textbf{Gamma(2,Z), $Z \sim Uniform(0.5,9)$$^{\tnote{\bf *}}$}} \\\cmidrule{3-5}\cmidrule{6-8}\cmidrule{9-11}
\textbf{Parameter}& \textbf{Method$^{\tnote{\bf **}}$} & \textbf{Estimate} & \textbf{MSE(MCSE)} & \textbf{Bias(MCSE)} & \textbf{Estimate} & \textbf{MSE(MCSE)} & \textbf{Bias(MCSE)} & \textbf{Estimate} & \textbf{MSE(MCSE)} & \textbf{Bias(MCSE)}\\
\midrule
$\beta_0=2$   & Adj-LM   & 1.8379     & 1.7823 (0.0714) & -0.1621 (0.0321) & 1.5212      & 2.2016 (0.1038) & -0.4788 (0.0467) & 0.203                            & 1.4037 (0.0655) & -0.032 (0.0311)  \\
          & Naive-LM & 2.6384     & 1.7339 (0.088)  & 0.6384 (0.0289)  & 2.593       & 2.1571 (0.1038) & 0.593 (0.0486)   & 0.7526                           & 1.3298 (0.0665) & -0.0206 (0.0322) \\
          & POI-SIMEX    & 2.1474     & 1.5458 (0.0736) & 0.1474 (0.0295)  & 1.6232      & 2.1037 (0.0965) & -0.3768 (0.0491) & 0.3627                           & 1.3972 (0.0661) & -0.0325 (0.0304) \\
          & True-LM  & 2.0019     & 1.3486 (0.0625) & 0.0019 (0.0291)  & 2.0483      & 1.3455 (0.0571) & 0.0483 (0.0415)  & 2.0050                            & 1.1637 (0.0737) & -0.0280 (0.0262)  \\
\midrule
$\beta_x=1$   & Adj-LM   & 1.0932     & 0.1747 (0.0108) & 0.0932 (0.0121)  & 1.0520       & 0.0088 (0.0006) & 0.0520 (0.0023)   & 1.7082                           & 0.0151 (0.0005) & 0.0764 (0.0017)  \\
          & Naive-LM & 0.6870      & 0.1512 (0.0048) & -0.3130 (0.0068)  & 0.9442      & 0.0081 (0.0004) & -0.0558 (0.0024) & 1.1763                           & 0.0156 (0.0006) & -0.0965 (0.0019) \\
          & POI-SIMEX    & 0.9343     & 0.1080 (0.0047)  & -0.0657 (0.0097) & 1.0422      & 0.0077 (0.0006) & 0.0422 (0.0023)  & 1.5431                           & 0.0110 (0.0004)  & 0.0512 (0.0017)  \\
          & True-LM  & 1.0081     & 0.0705 (0.0035) & 0.0081 (0.0073)  & 0.9985      & 0.0028 (0.0002) & -0.0015 (0.0015) & 1.0008                           & 0.0047 (0.0003) & 0.0014 (0.0017)  \\
\midrule
$\beta_z=0.5$ & Adj-LM   & 0.4953     & 0.0425 (0.0024) & -0.0047 (0.0060)  & 0.4984      & 0.0492 (0.0020)  & -0.0016 (0.0055) & -0.1663                          & 0.1068 (0.0044) & -0.1388 (0.0061) \\
          & Naive-LM & 0.4953     & 0.0403 (0.0024) & -0.0047 (0.0062) & 0.4970       & 0.0472 (0.0019) & -0.003 (0.0060)   & 0.6672                           & 0.1088 (0.0051) & 0.2007 (0.0063)  \\
          & POI-SIMEX    & 0.4953     & 0.0413 (0.0023) & -0.0047 (0.0062) & 0.4973      & 0.0495 (0.0022) & -0.0027 (0.0055) & 0.0962                           & 0.0910 (0.0044)  & -0.0894 (0.0059) \\
          & True-LM  & 0.4952     & 0.0383 (0.0020)  & -0.0048 (0.0060)  & 0.4928      & 0.0358 (0.0015) & -0.0072 (0.0055) & 0.4965                           & 0.0600 (0.0028)   & 0.0035 (0.0037) \\
\bottomrule
\end{tabular*}
\begin{tablenotes}
\item[$^{\rm *}$] In all settings, mean squared error (MSE), Bias and Monte Carlo standard error (MCSE) are based on the 1000 simulated datasets.
\item[$^{\rm **}$] Adj-LM:Li \textit{et al.}\cite{Lietal2004} proposed bias correction method; Naive-LM:Without bias correction (i.e., use observed covariate);POI-SIMEX: Poisson SIMEX bias correction; True-LM: use the true covariate from the simulation.
\end{tablenotes}
\end{sidewaystable}

\begin{sidewaystable}
\def\d{\hphantom{0}}
\caption{Appendix: Simulation results for misspecification:  $X_i \sim Gamma(a,b), W_i \lvert X_i \sim Neg-Binomial (5,P_i)$ where $P_i=5/(5+X_i)$, and sample size 100 \label{atab3}}

\begin{tabular*}{\textwidth}{@{\extracolsep\fill}lllllllllll@{}}
\toprule
& &\multicolumn{3}{@{}l}{\textbf{Gamma(1,2)$^{\tnote{\bf *}}$}} & \multicolumn{3}{@{}l}{\textbf{Gamma(1,10)$^{\tnote{\bf *}}$}} & \multicolumn{3}{@{}l}{\textbf{Gamma(2,Z), $Z \sim Uniform(0.5,9)$$^{\tnote{\bf *}}$}} \\\cmidrule{3-5}\cmidrule{6-8}\cmidrule{9-11}
\textbf{Parameter}& \textbf{Method$^{\tnote{\bf **}}$}  & \textbf{Estimate} & \textbf{MSE(MCSE)} & \textbf{Bias(MCSE)} & \textbf{Estimate} & \textbf{MSE(MCSE)} & \textbf{Bias(MCSE)} & \textbf{Estimate} & \textbf{MSE(MCSE)} & \textbf{Bias(MCSE)}\\
\midrule
$\beta_0=2$   & Adj-LM   & 4.0798     & 5.8195 (0.1451) & 2.0798 (0.0236)  & 12.242      & 111.5304 (1.1759) & 10.242 (0.0617)  & 2.1839                           & 2.7965 (0.1335) & 0.1839 (0.0501)  \\
          & Naive-LM & 4.0685     & 5.6941 (0.1389) & 2.0685 (0.0239)  & 12.2115     & 110.7669 (1.1811) & 10.2115 (0.0614) & 2.1617                           & 2.7111 (0.1292) & 0.1617 (0.0491)  \\
          & POI-SIMEX    & 4.0792     & 5.8068 (0.1441) & 2.0792 (0.0238)  & 12.2386     & 111.4459 (1.1728) & 10.2386 (0.0616) & 2.1810                            & 2.7928 (0.1313) & 0.181 (0.0495)   \\
          & True-LM  & 2.0091     & 1.2885 (0.0194) & 0.0091 (0.0311)  & 1.9987      & 1.4193 (0.0459)   & -0.0013 (0.0375) & 2.0033                           & 1.2667 (0.0442) & 0.0033 (0.0464)  \\
\midrule
$\beta_x=1$   & Adj-LM   & -0.0133    & 1.0393 (0.0117) & -1.0133 (0.0055) & -0.0574     & 1.1561 (0.0125)   & -1.0574 (0.0058) & -0.0387                          & 1.1006 (0.0144) & -1.0387 (0.0068) \\
          & Naive-LM & -0.0108    & 1.0308 (0.0096) & -1.0108 (0.0046) & -0.0498     & 1.1331 (0.0109)   & -1.0498 (0.0052) & -0.0339                          & 1.0869 (0.0131) & -1.0339 (0.0062) \\
          & POI-SIMEX    & -0.0131    & 1.0384 (0.0113) & -1.0131 (0.0053) & -0.0566     & 1.1539 (0.0125)   & -1.0566 (0.0058) & -0.0380                           & 1.0988 (0.0143) & -1.0380 (0.0067)  \\
          & True-LM  & 1.0064     & 0.0681 (0.0027) & 0.0064 (0.0051)  & 1.0018      & 0.0028 (0.0001)   & 0.0018 (0.0017)  & 0.9951                           & 0.0045 (0.0002) & -0.0049 (0.0018) \\
\midrule
$\beta_z=0.5$ & Adj-LM   & 0.4965     & 0.0427 (0.0013) & -0.0035 (0.0055) & 0.5027      & 0.1976 (0.0111)   & 0.0027 (0.0141)  & 2.4879                           & 4.0938 (0.0402) & 1.9879 (0.0098)  \\
          & Naive-LM & 0.4967     & 0.0426 (0.0013) & -0.0033 (0.0055) & 0.5027      & 0.1973 (0.0111)   & 0.0027 (0.0141)  & 2.4884                           & 4.0953 (0.0402) & 1.9884 (0.0098)  \\
          & POI-SIMEX    & 0.4965     & 0.0427 (0.0013) & -0.0035 (0.0055) & 0.5027      & 0.1977 (0.0111)   & 0.0027 (0.0141)  & 2.4879                           & 4.0936 (0.0402) & 1.9879 (0.0098)  \\
          & True-LM  & 0.4953     & 0.0373 (0.0013) & -0.0047 (0.0058) & 0.4987      & 0.0361 (0.0014)   & -0.0013 (0.0054) & 0.5039                           & 0.0589 (0.0028) & 0.0039 (0.0096) \\
\bottomrule
\end{tabular*}
\begin{tablenotes}
\item[$^{\rm *}$] In all settings, mean squared error (MSE), Bias and Monte Carlo standard error (MCSE) are based on the 1000 simulated datasets.
\item[$^{\rm **}$] Adj-LM:Li et al.\cite{Lietal2004} proposed bias correction method; Naive-LM:Without bias correction (i.e., use observed covariate);POI-SIMEX: Poisson SIMEX bias correction; True-LM: use the true covariate from the simulation.
\end{tablenotes}
\bigskip\bigskip  
\caption{Appendix: Simulation results for misspecification:  $X_i \sim Gamma(a,b), W_i \lvert X_i \sim Neg-Binomial (10,P_i)$ where $P_i=10/(10+X_i)$, and sample size 100 \label{atab4}}

\smallskip
\begin{tabular*}{\textwidth}{@{\extracolsep\fill}lllllllllll@{}}
\toprule
& &\multicolumn{3}{@{}l}{\textbf{Gamma(1,2)$^{\tnote{\bf *}}$}} & \multicolumn{3}{@{}l}{\textbf{Gamma(1,10)$^{\tnote{\bf *}}$}} & \multicolumn{3}{@{}l}{\textbf{Gamma(2,Z). $Z \sim Uniform(0.5,9)$$^{\tnote{\bf *}}$}} \\\cmidrule{3-5}\cmidrule{6-8}\cmidrule{9-11}
\textbf{Parameter}& \textbf{Method$^{\tnote{\bf **}}$} & \textbf{Estimate} & \textbf{MSE(MCSE)} & \textbf{Bias(MCSE)} & \textbf{Estimate} & \textbf{MSE(MCSE)} & \textbf{Bias(MCSE)} & \textbf{Estimate} & \textbf{MSE(MCSE)} & \textbf{Bias(MCSE)}\\
\midrule
$\beta_0=2$   & Adj-LM   & 3.9969     & 5.6742 (0.1712) & 1.9969 (0.0332)  & 12.2751     & 112.3795 (1.296)  & 10.2751 (0.0675) & 2.1397                           & 3 (0.1276)      & 0.1397 (0.0681)  \\
          & Naive-LM & 3.997      & 5.6196 (0.1669) & 1.997 (0.0319)   & 12.2575     & 111.896 (1.294)   & 10.2575 (0.0671) & 2.1278                           & 2.9236 (0.1227) & 0.1278 (0.0681)  \\
          & POI-SIMEX    & 3.9968     & 5.6720 (0.1729)  & 1.9968 (0.0337)  & 12.2731     & 112.3322 (1.2921) & 10.2731 (0.0675) & 2.1399                           & 3.0070 (0.1263)  & 0.1399 (0.0687)  \\
          & True-LM  & 1.9877     & 1.4421 (0.0890)  & -0.0123 (0.0197) & 2.0121      & 1.4706 (0.0772)   & 0.0121 (0.0458)  & 2.0098                           & 1.2022 (0.0485) & 0.0098 (0.0427)  \\
\midrule
$\beta_x=1$   & Adj-LM   & 0.0007     & 1.0021 (0.0042) & -0.9993 (0.0021) & -0.0245     & 1.0611 (0.0088)   & -1.0245 (0.0044) & -0.0129                          & 1.0329 (0.0038) & -1.0129 (0.0018) \\
          & Naive-LM & 0.0007     & 1.0015 (0.0039) & -0.9993 (0.0019) & -0.0223     & 1.0553 (0.0084)   & -1.0223 (0.0042) & -0.0117                          & 1.0297 (0.0035) & -1.0117 (0.0017) \\
          & POI-SIMEX    & 0.0007     & 1.0020 (0.0043)  & -0.9993 (0.0021) & -0.0242     & 1.0606 (0.009)    & -1.0242 (0.0045) & -0.0129                          & 1.0329 (0.0038) & -1.0129 (0.0018) \\
          & True-LM  & 1.0125     & 0.0694 (0.0040)  & 0.0125 (0.0075)  & 1.0004      & 0.0028 (0.0001)   & 0.0004 (0.0012)  & 1.0005                           & 0.0043 (0.0003) & 0.0005 (0.0024)  \\
\midrule
$\beta_z=0.5$ & Adj-LM   & 0.4990      & 0.0452 (0.0024) & -0.0010 (0.0061)  & 0.4736      & 0.1862 (0.0084)   & -0.0264 (0.0103) & 2.4864                           & 4.093 (0.0643)  & 1.9864 (0.0155)  \\
          & Naive-LM & 0.4990      & 0.0451 (0.0024) & -0.0010 (0.0060)   & 0.4736      & 0.1860 (0.0084)    & -0.0264 (0.0103) & 2.4868                           & 4.0943 (0.0641) & 1.9868 (0.0155)  \\
          & POI-SIMEX    & 0.4991     & 0.0452 (0.0024) & -0.0009 (0.0061) & 0.4737      & 0.1862 (0.0085)   & -0.0263 (0.0102) & 2.4863                           & 4.0926 (0.0642) & 1.9863 (0.0155)  \\
          & True-LM  & 0.4977     & 0.0380 (0.0021)  & -0.0023 (0.0048) & 0.4972      & 0.0394 (0.0022)   & -0.0028 (0.0045) & 0.4930                            & 0.0567 (0.0029) & -0.0070 (0.0096) \\
\bottomrule
\end{tabular*}
\begin{tablenotes}
\item[$^{\rm *}$] In all settings, mean squared error (MSE), Bias and Monte Carlo standard error (MCSE) are based on the 1000 simulated datasets.
\item[$^{\rm **}$] Adj-LM:Li et al.\cite{Lietal2004} proposed bias correction method; Naive-LM:Without bias correction (i.e., use observed covariate);POI-SIMEX: Poisson SIMEX bias correction; True-LM: use the true covariate from the simulation.
\end{tablenotes}
\end{sidewaystable}

\end{document}